\documentclass[aps,prb,reprint,twocolumn,showpacs,amsmath,amssymb]{revtex4-1}
\usepackage{amsfonts} % collection is a set of miscellaneous TeX fonts
\usepackage{amsmath,amssymb} % needed to make typesetting equations and symbols
\usepackage{bm}% bold math
\usepackage[usenames, dvipsnames]{color} % use colored tex
\usepackage{dcolumn}% Align table columns on decimal point
\usepackage{epic} % enhanced picture environment
\usepackage{epsfig} % handles eps files
\usepackage{eso-pic} % This package makes it easy to add some picture commands to every page.
\usepackage{eucal} % The package changes the font that is selected by \mathcal
\usepackage{graphicx} % for improved inclusion of graphics
\usepackage[breaklinks,colorlinks = true,linkcolor = red,urlcolor  = red,citecolor = red,anchorcolor = red,bookmarks=true]{hyperref}
\usepackage{mathrsfs}%\mathscr
\usepackage{amsthm}
\begin{document}
%-----------------------------------
\title{Quantum Destruction of Spiral Order in Two Dimensional Frustrated Magnets}
\author{Subhro Bhattacharjee}
\email{subhro@physics.utoronto.ca}

\affiliation{
Department of Physics,  Indian Institute of Science, Bangalore-560012, India.\\
Department of Physics, University of Toronto, Ontario M5S 1A7, Canada.\\
Department of Physics \& Astronomy, McMaster University, Hamilton, Ontario L8S 4M1, Canada.
}
\date{\today}
%-----------------------------------------------------------------------------------------------

\begin{abstract}
We study the fate of spin$-1/2$ spiral ordered two dimensional quantum antiferromagnets which are disordered by quantum fluctuations. A crucial role is played by the topological point defects of the spiral phase which are known to have a $Z_2$ character. Previous works established that a non-trivial quantum spin liquid phase results when the spiral is disordered {\em without} proliferating the $Z_2$ vortices. Here we show that when the spiral is disordered by proliferating and condensing these vortices, valence bond ordering occurs due to quantum Berry phase effects. We develop a general theory for this latter phase transition and apply it to a lattice model. This transition potentially provides a new example of a Landau-forbidden deconfined quantum critical point.
\end{abstract}
%-----------------------------------------------------------------------------------------------
\pacs{75.10.Jm, 71.27.+a, 75.30.Kz, 71.10Hf}
\maketitle
%-----------------------------------------------------------------------------------------------

%%%%%%%%%%%%%%%%%%%%%%%%%%%%%%%%%%%%%%%%%%%%%%%%%%%%%%%%%%%%%%%%%%%%%%%%%%%%%%%%%%%%%%%%%%%%%%%%
\section{Introduction}

In recent years we have learnt a great deal about possible quantum phase transitions out of Neel ordered phases of spin-$1/2$ quantum antiferromagnets in two spatial dimensions{\cite{2008_sachdev}}. Much of the work have focussed on circumstances where the Neel ordering is collinear ({\em i.e.}, where all the spins align along a common axis, but, alternate in direction). It is now known that such collinear Neel states can give way to paramagnetic valence bond solids (VBS), which break lattice translation symmetry but not spin rotation symmetry, through generic second order quantum phase transitions. Since the two phases (Neel and VBS) break incompatible symmetries of the Hamiltonian, a continuous transition between them, within the conventional paradigm of critical phenomena (known as the Landau-Ginzburg-Wilson paradigm \cite{1976_ma}),  is not allowed except at {\em fine-tuned} multi-critical points. Hence the above generic second order transition has been dubbed as {\em Landau-forbidden} \cite{2004_senthil}. The notion of {\it deconfined criticality} has been introduced  in the context of this and other similar phase transitions {\cite{2004_senthil}}. Possible continuous phase transitions between the collinear Neel phases and gapless spin liquids that do not break any symmetries have also been described {\cite{2005_ghaemi}}.

In contrast, quantum phase transitions out of the spiral ordered Neel phase ({\em i.e.}, with non-collinear spin patterns), that may occur on non-bipartite lattices, have not been as extensively studied. Early theoretical works{\cite{1994_chubukov,1991_read}} discussed continuous phase transitions from such a spiral state to a gapped $Z_2$ spin liquid with fractionalized spin-$1/2$ spinon (bosonic) excitations and associated non-magnetic $Z_2$ vortices (dubbed visons {\cite{2000_senthil}}). These studies have given rise to speculations that disordering spiral magnets naturally lead to deconfined spin liquids and the transition between the spiral phases and conventional paramagnets with valence bond order has remained unexplored. This is despite good theoretical motivations and some numerical evidence that exists for understanding such transitions as is reviewed below.

Quite generally one can view the nature of the transition out of the spiral from the perspective of its topological defects. These are the well-known point-like $Z_2$ vortices in two spatial dimensions {\cite{1984_kawamura}}. If the spiral is destroyed without proliferating these vortices a gapped $Z_2$ spin liquid with fractionalized bosonic spinon excitations emerges and the resulting transition is rather well understood-- the transition is described in terms of the condensation of the bosonic spinons, while the visons remain gapped throughout; the critical point belongs to $O(4)$ universality class \cite{1994_chubukov}. 

\begin{figure}
\centering
\includegraphics[scale=0.3]{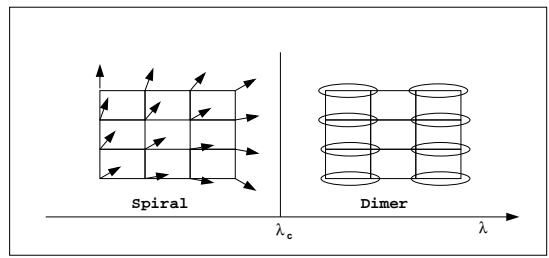}
\caption{Schematics of a generic continuous quantum phase transition between the spiral  and the valence bond paramagnet. The transition is brought about by the proliferation and condensation of the $Z_2$ vortices (refer to the text) achieved by tuning the parameter $\lambda$.}
\label{figlat}
\end{figure}

However, the more conventional transition out of the spiral is driven by proliferating the $Z_2$ vortices, such that both the spinons and the visons become critical at the transition. What is the nature of the resulting paramagnet and the associated transition? This is the fundamental question that we will formulate an answer to in this paper. We will argue that the quantum Berry phase effects lead to VBS order in such paramagnets and the associated transition may be generically second ordered. A schematic picture for a quantum phase transition between the spiral and the dimer is shown in Fig. \ref{figlat}. In this paper  we shall outline the general structure of the field theory for a generic continuous transition between the two phases and work out an example in context of a lattice model. We emphasize that the theory developed here can capture similar transitions on other lattices. Examples of such extensions are considered towards the end of this paper.  As pointed out earlier, since the spiral and the VBS break different incompatible symmetries of the Hamiltonian, a generic second order transition between them is an example of violation of the conventional Landau-Ginzburg-Wilson{\cite{1976_ma}} theory phase transitions.
 
The organization of the rest of the paper is as follows. We begin, in Sec. \ref{sec1}, by briefly reviewing some of the motivations that exists for studying quantum phase transitions out of spiral magnets. There are a number of numerical studies of frustrated spin-$1/2$ quantum magnets on a square lattice. Most pertinent to this paper are the exact diagonalization calculations on the ``$J_1-J_3$" model on a square lattice {\cite{1996_leung}}. Over a range of $J_3/J_1$ (see below) the system orders into a spiral pattern. In a different, though proximate, range a paramagnetic ground state apparently obtains. This later state is believed to have VBS order. Similarly, as we argue below, the same model on a rectangular lattice possibly has a direct quantum phase transition between the spiral and the VBS. 

To construct the critical theory, we find it convenient to start from the spiral phase and consider disordering it by proliferating its defects. To this end we discuss different aspects of the spiral order parameter in Sec. \ref{sec2}--the order parameter manifold and its topological defects, the $Z_2$ vortices. It is known that such vortices carry non-trivial Berry phases. The role of the vortices in the transition is effectively captured using the well-known spinon parametrization of the spiral order parameter. This is introduced in Sec. \ref{sec2b}. The lattice field theory consistent with different symmetries of the problem is then written down in Sec. \ref{sec3} and different limits of this theory are discussed. It is shown that in presence of the Berry phase, the paramagnet obtained by proliferation of the $Z_2$ vortices is indeed dimerized. The critical theory, containing both soft spinon and vison modes, is discussed in Sec. \ref{sec4}. The spinons and the visons see each other as source of $\pi$-flux (mutual semions). This arbitrarily long range statistical interaction between them is implemented through a mutual Ising Chern-Simmons interaction term. After a series of transformations, the final theory is written down in terms the spinons coupled to a $\mathbb{U}(1)$ gauge field. The gauge field is compact and the VBS order parameter is proportional to the monopole operator of the gauge field. The critical theory contains doubled-monopole operators where the gauge flux changes by $\pm 4\pi$. However, combining available analytical and numerical results, we argue that this doubled-monopoles are dangerously irrelevant right at the critical point, {\em i.e.}, while they are relevant in the paramagnetic phase, they are irrelevant at the critical point. Thus the gauge field is rendered non-compact at the critical point. This field theory, expected to describe the critical point, is an anisotropic version of  the $NCCP3$ field theories (see below).  The issue of extension of the present theory to other lattices is briefly taken up in Sec. \ref{other_lattices}. Finally we conclude with our remarks in Sec. \ref{sec5}. The details of various calculations are summarized in Appendices \ref{one_appen}-\ref{four_appen}.

%%%%%%%%%%%%%%%%%%%%%%%%%%%%%%%%%%%%%%%%%%%%%%%%%%%%%%%%%%%%%%%%%%%%%%%%%%%%%%%%%%%%%%%%%%%%%%%
\section{Lattice Models}
\label{sec1}

We are interested in spin-$1/2$ frustrated quantum antiferromagnets in two spatial dimensions described generic Hamiltonians
\begin{align}
\mathcal{H}=J\sum_{\langle ij\rangle}{\bf S}_i\cdot{\bf S}_j +\cdots,
\end{align}
where ${\bf S}_i$ is the spin-$1/2$ operator at the $i$th lattice site and $J>0$ denotes nearest neighbour antiferromagnetic exchanges. The ellipses denote other spin rotation invariant interactions that may be tuned to drive the phase transition between the spin ordered and the paramagnetic phases. Though our formulation is generically applicable to all cases of direct and continuous quantum phase transition between the spiral and the dimer, here, for concreteness, we consider a simple lattice model to illustrate our results. Extensions of these ideas to other relevant lattice systems are quite straight forward and indicated towards the end of the paper.

Consider spins on a {\it rectangular} lattice with first ($J_1$) and third neighbour ($J_3$) antiferromagnetic exchanges
\begin{align}
\label{05_rect_hamiltonian}
\nonumber
\mathcal{H}=&\sum_{{\bf r}}\left(J_1\ {\bf S_{r}\cdot S_{r + x}} + J_3\ {\bf S_{r}\cdot S}_{{\bf r +} 2{\bf x}}\right)\\
&+\lambda\sum_{{\bf r}}\left(J_1\ {\bf S_{r}\cdot S_{r + y}} + J_3\ {\bf S_{r}\cdot S}_{{\bf r +} 2{\bf y}}\right)
\end{align}
where ${\bf r}$ denotes the sites of the rectangular lattice. The couplings along $y$ direction are $\lambda$ (anisotropy factor) times those along the $x$ direction ($0\leq\lambda\leq 1$). The two tuneable parameters in the Hamiltonian are  $\lambda$ and $(J_3/J_1)$. The presence of the latter leads to frustration of exchange interactions. 

Various limits of this model are well known. For $\lambda=0$, one has decoupled spin chains with nearest and next nearest neighbour couplings, where, it is known that for $J_3/J_1<0.241$ there is power-law Neel order, while above this true long range VBS order is obtained {\cite{1992_nomura}}. On the other hand, $\lambda=1$ represents an isotropic square lattice with nearest and third nearest neighbour interactions. Numerical results {\cite{1996_leung,2004_capriotti}} suggest that this has at least three phases. For $J_3/J_1\lesssim 0.3$ the usual collinear Neel state obtains while for $J_3/J_1\gtrsim 0.7$ the ground state shows non-collinear spiral order. At intermediate values one gets a paramagnet which possibly breaks lattice symmetry, {\em i.e.} a VBS state. 

Imagine sitting in the spiral phase at $\lambda = 1$ (by choosing $J_3/J_1 > 0.7$). What happens if $\lambda$ is decreased towards zero? Clearly decreasing $\lambda$ increases quantum fluctuations so that the spiral order will be destroyed below some critical $\lambda_c$. For very small $\lambda$ the VBS order of the decoupled chains will persist as columnar dimer order with a 2-fold degenerate ground state. Could this VBS state persist all the way up to $\lambda_c$ ? If so could the resulting transition be second order? While the first question can only be answered by detailed numerical studies of this particular microscopic model in future we will formulate an answer to the second one in this paper. 

Having established the setting we now look at the spiral phase and consider disordering it through proliferation and condensation of its defects.

%%%%%%%%%%%%%%%%%%%%%%%%%%%%%%%%%%%%%%%%%%%%%%%%%%%%%%%%%%%%%%%%%%%%%%%%%%%%%%%%%%%%%%%%%%%%%%
\section{The Spiral Order Parameter and its topological defect}
\label{sec2}

Consider a magnetically ordered phase where the average spin orientations are coplanar. Such a spin pattern is characterized by  
\begin{align}
\label{05_spiral}
\langle{\bf S_{r}}\rangle\sim\left({\bf n_{1}}({\bf r}) \cos{({\bf Q\cdot r})} + {\bf n_{2}}({\bf r}) \sin{({\bf Q\cdot r)}}\right)\neq 0,
\end{align}
where, ${\bf Q}$ is the ordering vector which may be commensurate or incommensurate with the underlying lattice structure. For ${\bf Q}\neq n\pi$ ($n$ being an integer) the spin order is non-collinear and the forms a spiral pattern (as shown in Fig. \ref{figlat}). ${\bf n_{1}}({\bf r})$, ${\bf n_{2}}({\bf r})$ are two mutually orthogonal unit vectors, {\em i.e.}, 
\begin{align}
\label{eqnnconst}
{\boldsymbol n_{\alpha}}({\bf r})\cdot \boldsymbol n_{\beta}({\bf r})=\delta_{\alpha\beta}.
\end{align}
Here $\alpha,\beta=1,2$. The main difference between the collinear and spiral order parameter is that in the former only a single unit vector (${\bf n}_1$ or ${\bf n}_2$) is sufficient, while the latter requires both ${\bf n}_1$ and ${\bf n}_2$. The spiral order parameter is then given by an $\mathbb{SO}(3)$ matrix
\begin{align}
\mathcal{R}\equiv \left[{\bf n_{1}},\ {\bf n_{2}},\ {\bf n_1}\times{\bf n_2}\right].
\end{align} 

This difference in the order parameter manifold has several important consequences, the foremost being three spin-wave modes in the spiral phase compared to two in the collinear phase. Another feature, which is central to this paper, is the fact that topological defects of the spiral are distinct from that of the collinear Neel order. For the spiral phase, the order parameter has the symmetry of three-dimensional rotation group $\mathbb{SO}(3)$. The order parameter manifold is isomorphic to $S^3/Z_2$, which is an unit sphere in 4-dimensional space whose antipodal points are identified. Thus the order parameter space is {\it compact} and {\it doubly connected}\ {\cite{1985_tung}}.

In two spatial dimensions, this allows topologically stable point vortices that are characterized by a $Z_2$ quantum number{\cite{1984_kawamura}}
\begin{align}
\nonumber
 \Pi_1\left(S^3/Z_2\right)&=\mathbb{Z}_2,\\
\Pi_2\left(S^3/Z_2\right)&=0,
\end{align} 
where $\Pi_1$ and $\Pi_2$ denotes the first and second homotopy groups respectively {\cite{1979_mermin}}. These $Z_2$ vortices are distinct from the O(2) point vortices in a familiar two-dimensional $XY$-model. In case of the $Z_2$ vortices, they can have a winding number which is either 0 or 1 (i.e. the winding number is defined modulo 2). This is a reflection of the fact that there are only two classes of closed path in the $S^3/Z_2$ manifold and all closed paths can grouped in one of these two classes. Hence, the $Z_2$ vortices are characterized by an Ising quantum number. Two such vortices can annihilate each other.
 
In the ordered phase the energy of a single vortex diverges logarithmically with the system size and free vortices are absent. However the vortices play a crucial role when the spiral order is destroyed by condensing them. To this end it will be extremely useful to set up an effective description of the spiral and proximate paramagnetic phases that captures easily the role of the $Z_2$ vortices. The formulation in terms of ${\bf n}_{1,2}$ along with the constraints (Eq. \ref{eqnnconst}) is not convenient for this purpose. Instead it is useful to introduce an alternative parameterization of the spiral order parameter. 

%----------
\subsection{The spinon parametrization of the spiral order parameter and the Ising gauge fields}
\label{sec2b}

We introduce the well-known redundant description of the order parameter in terms of the spinon variables. Specifically we write {\cite{1994_chubukov,1990_angelucci}}
\begin{align}
\label{05_n_z}
{\bf n^{+}}={\bf n_{1}}+ \imath{\bf n_{2}}=\epsilon_{\alpha\beta}z_{\beta}{\boldsymbol\sigma}_{\alpha\gamma}z_{\gamma},
\end{align}
where ${\boldsymbol\sigma}$s are the three Pauli matrices and $\epsilon_{\alpha\beta}$ is the two dimensional completely antisymmetric matrix $(\epsilon_{12}=-\epsilon_{21}=-1,\epsilon_{11}=\epsilon_{22}=0)$. ${\bf z}=(z_1,z_2)$ is the  two component complex spinon field with unit norm, {\em i.e.},
\begin{align}
z^*_\alpha z_\alpha=1.
\end{align}
 The spinons are thus bosons that transform as spin-1/2 under spin rotations. We note that the spinons defined here are related to the Caley-Klein parameters (which in turn are related to the Euler angles) used in describing the motion of a classical rigid body {\cite{2001_goldstein}}.

At this point it is useful to consider the global symmetries of the effective action written in terms of the spinon fields. The action must be invariant under global SU(2) spin rotation under which the spinons transform as
\begin{align}
\label{spin_sym_trans}
z_\alpha\rightarrow V_{\alpha\beta}z_\beta,
\end{align}
where $V_{\alpha\beta}$ is an $\mathbb{SU}(2)$ matrix. In addition there are several other discrete symmetries of the Hamiltonian- various lattice transformations and time reversal. For a rectangular lattice, the transformations of the spinon under these symmetries are given by :
\begin{align}
\nonumber
\mathcal{T}_{\bf a} 	  &: z_\alpha \rightarrow	e^{-\imath {\bf Q}\cdot{\bf a}/2}z_\alpha,\\
\nonumber
\label{ref_sym_trans}
\mathcal{R}_s,\mathcal{I} &: z_\alpha \rightarrow	\sigma^y_{\alpha\beta}z^*_\beta,\\
T			  &: z_\alpha \rightarrow	iz^*_\alpha,
\end{align}
Where $\mathcal{T}_a$ represents unit translation along {\bf a}, $\mathcal{R}_s$ denotes reflection about an axis of symmetry of the lattice, $\mathcal{I}$ denotes inversion of the lattice and $T$ denotes time reversal. The Translation symmetry implies that the continuum theory has an extra $\mathbb{U}(1)$ symmetry along with the spin $\mathbb{SU}(2)$ and other discrete symmetries.

The spinon parameterization is {\it two-to-one} and there is a  discrete $\mathbb{Z}_2$ gauge symmetry corresponding to the change of sign of the ${\bf z}$ fields {\em independently} at each site
\begin{align}
{\bf z}({\bf r})\rightarrow -{\bf z}({\bf r})
\end{align}
which leaves the order parameter invariant. This reiterates the fact, in terms of the spinons, that the order parameter manifold is $S^3/Z_2$. 

Around a vortex the order parameter fields ${\bf n}_\alpha$ are single valued. However the spinon fields, $z_\alpha$, changes sign. Thus the spinon wave function changes its sign on going around a vortex. In other words the spinons and the $Z_2$ vortices see each other as sources of $\pi$ flux. 
 
 Following Lammert {\em et al.}{\cite{1993_lammert}}, a fruitful description of the vortices is now achieved by introducing an Ising gauge field $\sigma_{ij}=\pm 1$, minimally coupled to the spinons, on the links of the direct lattice.  The $Z_2$-vortices are then associated with the magnetic flux of a $Z_2$ gauge field
\begin{align}
\mathcal{F}_\Box=\prod_{\Box}\sigma_{ij},
\end{align}
where the product is taken over the links of the plaquette of the Ising gauge field. $\mathcal{F}_\Box=-1(+1)$  indicates the presence(absence) of a vison inside the plaquette. It is important to note {\cite{2003_sachdev_2}} that visons are well defined excitations even in the paramagnetic phase. Here, the spinon fields fluctuate wildly and so does their corresponding paths in the order parameter space. However, for spinons describing a closed loop around a vison, the paths necessarily end at diametrically opposite points in order parameter manifold ($S^3/Z_2$).

%----------
\subsubsection*{The Berry Phase of the $Z_2$ vortices}

The above classical picture must be augmented with the correct quantum Berry phase term. Semi-classical analysis {\cite{1989_dombre,1993_diptiman}} for the spiral phase shows that the non-trivial Berry phases are solely associated with the vortices and are given by 
\begin{align}
\label{berry_phase}
e^{\mathcal{S}_B}=\exp{\left[\frac{i\pi}{2}\sum_{i,j=i+\tau}(1-\sigma_{ij})\right]}=\prod_{i,j=i+\tau}\sigma_{ij},
\end{align}
where $\sigma_{i,i+\tau}$ are the $Z_2$-gauge fields on the time-like links of the $(2+1)$D space-time lattice. A different study {\cite{2000_sachdev}}, starting from the spin disordered phase, recovers the same Berry phase term. Eq. \ref{berry_phase} is also, not coincidentally, the $Z_2$-Polyakov loop term obtained in the analysis of the {\it quantum dimer models} {\cite{2001_moessner}}.

It is useful to introduce{\cite{2000_senthil}} a new set of Ising gauge field $\mu^{ext}_{ab}=\pm 1$ (not to be confused with the $\mu_{ab}$ introduced later. We have used same notations as in Ref. \onlinecite{2000_senthil} for easy reference) on the links of $(ab)$ dual (2+1) dimensional lattice such that 
\begin{align}
\prod_\Box\mu^{ext}_{ab} =-1
\end{align}
for all the space-like plaquettes of the dual lattice. Then one can write the Berry phase in Eq. \ref{berry_phase} as
\begin{align}
\label{final_berry}
\mathcal{S}_B= \frac{\imath \pi}{4} \sum_{\langle ij\rangle}(1-\sigma_{ij})(1-\prod_{\Box}\mu^{ext}_{ab}),
\end{align}
where the links $(ij)$ and $(ab)$ belong to the direct and the dual lattices respectively such that the dual plaquette is pierced by the link $(ij)$.

%%%%%%%%%%%%%%%%%%%%%%%%%%%%%%%%%%%%%%%%%%%%%%%%%%%%%%%%%%%%%%%%%%%
\section{Lattice field Theory}
\label{sec3}

Various global symmetries and the $\mathbb{Z}_2$ gauge structure now fully determine the effective action which is invariant under the transformation group 
\begin{align}
\left[\mathbb{SU}(2)\times \mathbb{U}(1)\right]_{global}\times\left[\mathbb{Z}_2\right]_{gauge}
\end{align}
plus discrete lattice symmetries and time reversal (see Eq. \ref{ref_sym_trans}). The minimal imaginary time Ginzburg-Landau action  in $(2+1)$ space-time lattice, consistent with these symmetries, is 
\begin{align}
\label{05_first_action}
\mathcal{S}=\mathcal{S}_z+\mathcal{S}_B,
\end{align}
where
\begin{align}
\label{spinon_action}
\nonumber
\mathcal{S}_z=-t_s\sum_{\langle ij\rangle}\sigma_{ij}\left(z_{i}^\dagger\cdot z_{j} + h.c\right) -r\sum_{\langle ij\rangle}\left(z_i^\dagger\cdot z_j-z_j^\dagger\cdot z_i\right)^2 \\
\end{align}
and $\mathcal{S}_B$ is the Berry phase contribution given by Eq . \ref{berry_phase}. Eq. \ref{05_first_action} may be derived more formally using Hubbard-Stronovich transformation as shown in Appendix \ref{one_appen}. Integrating out the higher energy spinons generate several terms allowed by the symmetry, the foremost being the $Z_2$-{\it Maxwell} term which gives kinetic energy to the gauge fields
\begin{align}
\mathcal{S}_\sigma=-\sum_{P}K_P\prod_{\Box} \sigma_{ij},
\label{gauge_dynamics}
\end{align}
where the sum is taken over the plaquettes of the space-time lattice and $K_P$ depends on the plaquette orientation. Thus the effective low energy action is given by
\begin{align}
\mathcal{S}_{eff}=\mathcal{S}_z + \mathcal{S}_{\sigma} + \mathcal{S}_{B},
\label{05_final_eff_action}
\end{align}
where $\mathcal{S}_z,\mathcal{S}_{\sigma}$ and $\mathcal{S}_{B}$ are given by Eqs. \ref{berry_phase}-\ref{gauge_dynamics}. This action given by Eq. \ref{05_final_eff_action} is powerful enough to capture the large part of the phase diagram including the spiral, dimer and the $Z_2$ spin liquid phases and the associated transitions. However, in this paper we shall focus exclusively of the possibility of the direct second order transition between the spiral and the dimer phases.

%---------------------------------------------
\subsection*{The VBS Phase}

The spiral phase is obtained by condensing the spinons, {\em i.e.} $\langle z_\alpha\rangle\neq 0$ (hence $\langle {\bf n}\rangle\neq 0$). On the other hand, deep inside the paramagnet the spinon excitations are gapped and can be integrated out from Eq. \ref{05_final_eff_action} to yield the effective action
\begin{align}
\label{05_action_vbs}
\mathcal{S}'=\mathcal{S}_\sigma + \mathcal{S}_B.
\end{align}
This is the action for the {\it Odd Ising gauge theory} (Odd-IGT) {\cite{1971_wegner}}  which have been studied extensively in context of {\it quantum dimer models}{\cite{2001_moessner}}. In $(2+1)$ dimensions, Odd-IGT is fruitfully studied using Ising duality {\cite{1979_kogut,1980_savit}} connecting the Odd-IGT to the fully frustrated transverse field Ising model on the dual lattice {\cite{2000_senthil}}. The dual action is given by  
\begin{align}
\label{05_fftfim_ham}
\mathcal{H}_{FFTFIM}=\sum_{ab}\tilde{K}_{ab}\rho_a^z\mu^{ext}_{ab}\rho_b^z-\Gamma\sum_{ab}\rho_a^x.
\end{align}
Here $\rho^\alpha$ are dual Ising spins that occupy the sites of the dual lattice. The Berry phase  (Eq. {\ref{berry_phase}}) imposes the condition of full frustration in this dual model through $\mu_{ab}^{ext}$ as defined in Eq. \ref{final_berry}. We note again that these fields obey the constraint $\prod_{\Box}\mu_{ab}^{ext}=-1$ over all space-like dual plaquettes {\cite{2000_senthil}}. Without loss of generality we take the dual exchange couplings $\tilde{K}_{ab}>0$.

As derived in Appendix \ref{two_appen}, the vison creation operator, $\psi^\dagger_{vison}(a)$, is proportional to $\rho_a^z$, {\em i.e.},
\begin{align}
\psi_{vison}^\dagger (a)\sim \rho_a^z.
\end{align}
We note that the vison resides in the plaquettes of the direct lattice which is equivalent to the sites $(a)$ of the dual lattice. Also the VBS order parameter,  $\Psi_{VBS}(ij)$ is proportional to $\langle \rho^z_a\rho^z_b\rangle$, {\em i.e.}
\begin{align}
\Psi_{VBS}(ij)={\bf S}_i\cdot {\bf S_j}\sim \langle \rho^z_a\rho^z_b\rangle,
\end{align}
where, as described in Appendix \ref{two_appen}, the link $(ij)$ on the direct lattice crosses the link $(ab)$ on the dual lattice. We note that The VBS order parameter is bilinear in vison operators and is gauge invariant as one expects. 

The dual Ising spins undergo an ordering transition and the ordered phase is characterized by 
\begin{align}
\label{vison_cond}
\langle \rho^z_a\rangle\neq 0.
\end{align}
However due to the frustration, the ordered phase breaks lattice translation symmetry {\cite{2000_senthil}}. Eq. \ref{vison_cond} indicates that the visons proliferate and condense in this phase. The VBS order parameter gains a non-zero expectation value 
\begin{align}
\Psi_{VBS}\neq 0
\end{align}
and reflects the broken lattice translation symmetry. Thus the vison condensate indeed describes a dimerized paramagnet {\cite{1991_jalabert}}.

%-------------------------------------
\subsubsection*{The critical vison modes}

\begin{figure}
\centering
\includegraphics[scale=0.5]{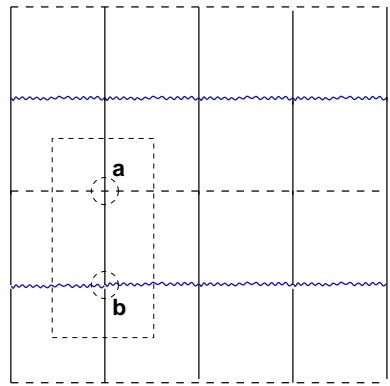}
\caption{The Fully frustrated model on the rectangular lattice: The black bonds (both solid and dashed) are ferromagnetic and the blue (wavy) bonds are antiferromagnetic. The gauge is chosen in such a way that around each plaquette that $\prod_{\Box}\mu_{ab}^{ext}=-1$ is satisfied. Further the magnitude of the solid black bonds are taken to be $\eta$ while those of the dashed black bonds and the blue bonds are taken to be 1, where $\eta > 1$. The unit-cell, as shown in the figure, contains two sites.}
\label{dual_fig}
\end{figure}

The momenta of the vison modes that condenses describes the VBS ordering pattern near the transition. Following Blanckstein {\em et al.} \cite{1984_blanckstein}, these modes can be obtained from a soft mode analysis of Eq. \ref{05_fftfim_ham}. In our case, the dual Ising model is described on the dual rectangular lattice. There is a gauge redundancy in choosing $\mu_{ab}^{ext}$ in Eq. \ref{05_fftfim_ham} and maintaining the constraint $\prod_{\Box}\mu_{ab}^{ext}=-1$ over all space-like plaquettes (see discussion following Eq. \ref{05_fftfim_ham}). In the rest of this calculation we shall assume, without any loss of generality, a specific gauge to identify the soft modes {\cite{2001_lannert}}.  This is shown in Fig. \ref{dual_fig}.
There are two soft modes: $\Lambda_0$ and $\Lambda_\pi$ at wave-vectors $(0,0)$ and $(\pi,0)$ respectively. 
\begin{align}
\Lambda_0=\left[\begin{array}{c}
1\\
\frac{1+\sqrt{1+\eta^2}}{\eta}\\
\end{array}
\right],\ \
\Lambda_\pi=\left[\begin{array}{c}
\frac{1+\sqrt{1+\eta^2}}{\eta}\\
1\\
\end{array}
\right]e^{\imath\pi x}
\end{align}
(Note that the modes are un-normalized. Also see Fig. \ref{dual_fig} for definition of the dual anisotropy factor $\eta$.)

The low energy vison modes near the transition are linear combination of $\Lambda_0$ and $\Lambda_\pi$: 
\begin{align}
\Phi({\bf r},t)=\Psi_0({\bf r},t)\Lambda_0+\Psi_\pi({\bf r},t)\Lambda_\pi,
\end{align}
where $\Psi_0({\bf r},t)$ and $\Psi_\pi({\bf r},t)$ are the two complex amplitudes. The effective Ginzburg-Landau action for the soft vison modes can be constructed by considering transformations of the two amplitudes under various symmetries of the Hamiltonian. Due to the gauge redundancy in choosing $\mu^{ext}_{ab}$, the transformations we need to consider are the projective symmetry transformations (PSG) of the soft vison modes {\cite{2002_wen}}.  It is useful to define a single complex vison mode
\begin{align}
\Psi=\Psi_0+i\ \Psi_\pi=\vert\Psi\vert e^{i\phi}
\end{align}
The Projective symmetry transformations for the soft vison modes are given by:
\begin{align}
\nonumber
\mathcal{T}_{\bf x},R_y 		:&	\Psi	\rightarrow 	\Psi^*,\\
\nonumber
\mathcal{T}_{\bf y}		:&	\Psi	\rightarrow	-\imath\Psi^*,\\
\mathcal{R}_\pi,\mathcal{I},R_x	:&	\Psi	\rightarrow	\Psi.
\end{align}
Here $\mathcal{T}_{\bf x}$ and $\mathcal{T}_{\bf y}$ are unit translation along ${\bf x}$ and ${\bf y}$ respectively, $\mathcal{R}_\pi$ is rotation by $\pi$ about a lattice point, $R_x, R_y$ are reflections about $x$ and $y$ axes respectively and $\mathcal{I}$ is Lattice Inversion. The lowest order action allowed by these transformations is given by
\begin{align}
\label{rect_soft_spin_LGW}
\mathcal{S}_{v}=-t_v\sum_{\langle a,b\rangle} \cos{(\phi_{a}-\phi_{b})} - \Gamma_v\sum_{a}\cos{(4\phi_{a})}.
\end{align}
The modes transform as {\it ``XY spin''} under different symmetries of the Hamiltonian up to $4^{th}$ order. As outlined before, in terms of the original spins, the ordered phase of this theory represents the VBS and  $e^{i\phi}$ is identified with the vison creation operator.

Having outlined the critical spinon and the vison modes we now proceed to construct the critical field theory.

%%%%%%%%%%%%%%%%%%%%%%%%%%%%%%%%%%%%%%%%%%%%%%%%%%%%%%%%%%%%%%%%%%%%
\section{The Critical Field Theory}
\label{sec4}
 
At the critical point both visons and spinons are gapless. Further, they see each other as sources of $\pi$ flux (mutual {\it semions}). Hence, there is an arbitrarily long range statistical interaction between them. Such interactions may be effectively implemented by introducing two Ising gauge fields ${\sigma_{ij}}({\mu_{ab}})$, on the links of the direct(dual) lattice, coupled by an Ising Chern-Simmons term {\cite{2000_senthil}} :  
\begin{align}
\label{chern_simmonsterm}
\mathcal{S}_{CS}=\frac{i\pi}{4}\sum_{\langle ab\rangle} (1-\prod_{\Box}\sigma_{ij})(1-\mu_{ab}).
\end{align} 
Here we note that this mutual Ising Chern-Simmons coupling is different from the mutual $U(1)$ Chern-Simmons term used for similar problems in Ref. {\onlinecite{2009_xu}}. We also take this opportunity to distinguish between the approach of the present paper and that of Ref. \onlinecite{2009_xu}, which is based on the mutual $U(1)$ Chern-Simmons theory. In particular, we focus on the latter's limitation in capturing the possible direct second order transition between the spiral and the dimer. In their work Xu {\em et al.} introduced the critical spinons and the visons like this work. However in their continuum theory they use two $U(1)$ gauge fields coupled via a Chern-Simmons term to implement the mutual semionic statistics between the spinons and the visons. This is valid, as the authors acknowledge, when the number of spinon and vison excitations are small\cite{2008_kou}. However near the spiral to dimer transitions there is a proliferation of low energy spinon and soft vison excitations and the $U(1)$ theory breaks down. On the other hand the mutual $U(1)$ Chern-Simmons theory successfully captures the transition between the spiral and a $Z_2$ spin liquid or collinear Neel phase. Indeed, the phase diagram obtained in that formalism shows that the spiral and the dimer phases are always separated, either by a $Z_2$ spin liquid or a collinear Neel phase. A direct transition between the two requires fine tuning to a multi-critical point and hence is not generic. On the other hand the present formalism can tackle the situation as we outline below. 

Following the short digression we now return to the discussion of the critical theory. The critical Landau-Ginzburg action is 
\begin{align}
\label{the_theory}
\mathcal{S}_{C}=\mathcal{S}_v+\mathcal{S}_z +\mathcal{S}_{CS},
\end{align}
where the different terms are given by Eqs. \ref{spinon_action}, \ref{rect_soft_spin_LGW} and \ref{chern_simmonsterm} respectively. The long range interaction between the spinons and the visons, as encoded in $\mathcal{S}_{CS}$, makes the analysis of this field theory difficult and a series of transformations are required to cast the theory in a useful form. To start with, we neglect the effect of the terms with coefficient $r$ (in Eq. \ref{spinon_action}) and $\Gamma_v$ (in Eq. \ref{rect_soft_spin_LGW}) and consider their effects later.

Within Villain approximation {\cite{2000_senthil,1977_villain}}, the first term in the vison action (Eq. \ref{rect_soft_spin_LGW}) can be written as (details of the calculations are given in Appendix \ref{three_appen}) 
\begin{align}
\mathcal{S}_v=\sum_{\langle ab\rangle} \frac{J_{ab}^2}{2t_v},
\end{align}
where $J_{ab}$ is the integer valued vison current defined on the links of the dual space-time lattice. The constraint on $J_{ab}$, as derived in Appendix \ref{three_appen}, are   
\begin{align}
\label{05_current_condition}
\boldsymbol{\nabla}\cdot {\bf J}=0
\end{align}
and
\begin{align}
(-1)^{J_{ab}}=\prod_{\Box}\sigma_{ij}
\end{align}
The first constraint is the continuity equation for the vison current reflects the fact that for $\Gamma_v=0$, in Eq. \ref{rect_soft_spin_LGW}, the vison number is conserved. The second constraint, on the other hand is the restatement of the mutual semionic statistics between the spinons and the visons. The zero divergence constraint is satisfied by defining an integer valued vector field ${\bf a}$ on the links of the direct lattice such that 
\begin{align}
\label{curl_condition}
{\bf J}=\boldsymbol{\nabla}\times {\bf a}.
\end{align}
The second constraint then becomes
\begin{align}
\sigma_{ij}=e^{i\pi a_{ij}}.
\label{05_mu_sigma_condition}
\end{align}
We now write 
\begin{align}
a_{ij}=2b_{ij}+s_{ij}
\end{align}
where $b_{ij}$ is an integer field and $s_{ij}=0(1)$ when $a_{ij}$ even(odd). Eq. \ref{05_mu_sigma_condition} then gives $\sigma_{ij}=1-2s_{ij}$. 

The condition on $a_{ij}$ to be an integer may be implemented by applying a {\it soft} potential:
\begin{align}
V_{soft}=-g\sum_{ij}\cos{(2\pi b_{ij})}=-g\sum_{ij}\sigma_{ij}\cos{(\pi a_{ij})}
\end{align}
where $g>0$. At this stage it is useful to break the 2 complex spinon fields, $z_1,z_2$ into a 4-component real vector field: 
\begin{align}
\nonumber
z_{1}&=\nu_1 +i\nu_2,\\
z_{2}&=\nu_3 + i\nu_4.
\end{align}
Further, rescaling the gauge potential $a_{ij}\rightarrow a_{ij}/\pi$ and choosing the transverse gauge 
\begin{align}
\boldsymbol{\nabla}\cdot{\bf a}=0
\end{align}
by defining a scalar field $\zeta$ on the direct lattice such that 
\begin{align}
a_{ij}\rightarrow a_{ij}+\Delta\zeta_{ij}
\end{align}
we have
\begin{align}
\nonumber
\mathcal{S}_C=&&-\sum_{ij}\sigma_{ij}\left[t_s{ {\boldsymbol\nu}_i\cdot{\boldsymbol{\nu}}_j}+g\cos{(a_{ij}+(\zeta_{i}-\zeta_j)})\right]\\
&&+\frac{1}{2t_v\pi^2}\sum_{ij}(\nabla_{ij}\times a_{ij})^2,
\end{align}
where ${\boldsymbol{\nu}}=(\nu_1,\nu_2,\nu_3,\nu_4)$. Summing over $\sigma_{ij}$s and collecting the relevant coupling terms give the critical action to the lowest order
\begin{align}
\mathcal{S}_C=-\sum_{ij}\left[t\left(e^{ia_{ij}}{ \boldsymbol{\chi}_i\cdot\boldsymbol{\chi}^*_j}+ h.c\right)- \frac{\left(\nabla_{ij}\times a_{ij}\right)^2}{2t_v}\right],
\end{align}
where we have introduced complex 4-component vectors
\begin{align}
\chi_{i\alpha}=\nu_{i\alpha}e^{i\zeta_{i}}.
\end{align}
The constraint over the ${\bf \chi}$ fields being that they are uni-modular and parallel, {\em i.e.}, 
\begin{align}
\nonumber
\chi^*_\alpha\chi_\alpha&=1\\
\chi^*_\alpha\chi_\beta-\chi^*_\beta\chi_\alpha&=0.
\end{align}
This is implemented by incorporating a soft potential: 
\begin{align}
V'_{soft}=\eta_0(1-(\chi_i^*)^2(\chi_i)^2)
\label{chi_potential}
\end{align}
where $\eta_0 > 0$. The continuum limit for this critical theory may now be written using a {\it soft-spin} description. 
\begin{align}
\label{05_contd_fth}
\nonumber
\mathcal{S}_{eff}=\int d^2x\ d\tau\ &&\left[\vert\left(\partial_\mu-ia_\mu\right){\boldsymbol\chi}\vert^2 + p\vert{\boldsymbol\chi}\vert^2+u\left(\vert{\boldsymbol\chi}\vert^2\right)^2\right.\\
&&\left.-\eta_0({\boldsymbol\chi}^*)^2({\boldsymbol\chi})^2+\frac{1}{e^2}\left(\nabla\times {\bf a}\right)^2\right]
\end{align}

So far we have neglected the effect of $r$ coupling present in the microscopic model (Eq. \ref{spinon_action}). Without this term the symmetry is enlarged from the original microscopic $\mathbb{SU}(2)\times\mathbb{U}(1)$ to $\mathbb{SO}(4)$. Hence we expect that on taking $r\neq 0$ in the microscopic model , terms allowed by the microscopic symmetry would be generated on coarse graining  and reduce this enlarged symmetry back to the microscopic one. Such a generic term can be found starting from the universal covering group $\mathbb{SU}(4)$ of uni-modular 4-component complex vectors. 

The $\mathbb{SU}(4)$ is generated by fifteen $4\times 4$ matrices $\{\tau_a,\mu_a,\tau_a\mu_b\}$ where
\begin{align}
\tau_a=\left[\begin{array}{cc}
\sigma_a & 0\\
0 & \sigma_a\\
\end{array}\right], \ \
\mu_a=\left[\begin{array}{cc}
0 & \sigma_a\\
\sigma_a & 0\\
\end{array}\right]
\end{align}
where $\sigma^a$ are the usual Pauli matrices and $\tau^a\mu^b$ implies direct product. Out of these, the 6 purely imaginary antisymmetric ones that generate $\mathbb{SO}(4)$ are  
\begin{align}
\left\{\tau_y,\ \ \mu_y,\ \ \tau_x\mu_y,\ \ \tau_y\mu_x,\ \ \tau_y\mu_z,\ \ \tau_z\mu_y\right\}.
\end{align}
Since $\mathbb{SO}(4)\equiv \mathbb{SU}(2)\times \mathbb{SU}(2)$ these 6 generators can be broken into 2 mutually commuting sets:
\begin{align}
\left\{\mu_y,\ \ \tau_y\mu_z,\ \ \tau_y\mu_x\right\},\ \ \left\{\tau_y,\ \ \tau_x\mu_y,\ \ \tau_z\mu_y\right\} 
\label{eq_su2Xsu2}
\end{align}
The $\mathbb{SU}(2)$ group generated by the first set of generators represent spin rotations (see Appendix \ref{four_appen}). Now we can choose a generic term that breaks the second $\mathbb{SU}(2)$ to $\mathbb{U}(1)$. This residual $\mathbb{U}(1)$ can then be identified with the global $\mathbb{U}(1)$ symmetry under lattice translation (refer to our previous discussion). This is easily done by considering the most generic term at the lowest order allowed by various point group symmetries. The transformations are given in Appendix \ref{four_appen} with the result being 
\begin{align}
\left({\bf \boldsymbol\chi^*\cdot\tau_y\cdot\boldsymbol\chi}\right)^2
\end{align}
 (the term ${\bf \boldsymbol\chi^*\cdot\tau_y\cdot\boldsymbol\chi}$ is forbidden by time reversal). So the final continuum field theory is 
\begin{align}
\label{05_final_continum_field_theory}
\nonumber
\mathcal{S}_{eff}=\int d^2xd\tau &\left[\vert\left(\partial_\mu-ia_\mu\right){\bf \boldsymbol\chi}\vert^2 + p\vert{\bf \boldsymbol\chi}\vert^2+u\left(\vert{\bf\boldsymbol\chi}\vert^2\right)^2\right.\\
\nonumber
&\left.-\eta_0({\bf \boldsymbol\chi}^*)^2({\bf \boldsymbol\chi})^2+\gamma\left({\bf \boldsymbol\chi^*\tau_y \boldsymbol\chi}\right)^2\right.\\
&\left. -\frac{1}{e^2}\left(\nabla\times {\bf a}\right)^2\right].
\end{align}
Thus the final field theory is described in terms of 4 component complex matter fields coupled to a non-compact $\mathbb{U}(1)$ gauge field.

%----------------------
\paragraph*{Fate of Doubled Instantons:}

Now consider the role of $\Gamma_v$ (in Eq. \ref{rect_soft_spin_LGW}), which introduces the $4$-fold anisotropy. For $\Gamma_v=0$, from Eq. \ref{rect_soft_spin_LGW}, the (vison)number operators conjugate to the $\phi$ fields are conserved (Eq. \ref{05_current_condition}).  This is equivalent to the flux conservation of the $\mathbb{U}(1)$ gauge field $a_\mu$ (see Eq. \ref{curl_condition}). Finite $\Gamma_v$ destroys this conservation. Remembering that $e^{i\phi}$ is a vison creation operator, we see that the $\Gamma_v$  term allows the simultaneous appearance/disappearance of 4 visons or change of gauge flux by $\pm 4\pi$,  {\em i.e.}, a {\it doubled instanton} operator {\cite{1987_polyakov}}.

In Eq. \ref{05_final_continum_field_theory}, the condensation of the spinons lead to spiral ordering. Once they condense, the gauge field dynamics is gapped through the Anderson-Higg's mechanism and the instantons are suppressed. On the other hand the instantons are relevant in the paramagnetic phase and their condensation lead to VBS order. Thus this field theory captures the right limits and suggests that there may be a direct transition between the spiral (Higgs phase) and the dimer (confined phase) {\cite{1979_fradkin}}. 

This direct transition may be continuous only if the doubled instantons are irrelevant at the critical point. Presently, accurate estimates of the scaling dimension of this doubled instanton operator is missing. While these may be obtained numerically in future, here we make a crude estimate. Large-$N$, RPA treatment of the gauge fluctuations {\cite{1990_murthy}}, suggest that the scaling dimensions of instanton of charge $q$ is proportional to $q^2N$ {\cite{2007_grover}}. Recent numerical studies {\cite{2007_sandvik}} on the isotropic $NCCP^1$ model find that a single instanton has a scaling dimension of $0.63$. Combining these, we find that the scaling dimension of the doubled$(q=2)$ instanton operator $(\Delta)$ is $\Delta=\frac{4}{2}\times 2^2\times (0.63)\approx 5.04>3$. Hence this naive estimate suggests that doubled instantons are irrelevant at the critical point and so the $\mathbb{U}(1)$ gauge flux is conserved right at the critical point. This emerging $\mathbb{U}(1)$ symmetry, absent in the microscopic model, is typical to {\it deconfined quantum critical points}. An extensive characterization of such critical points is given in Ref. {\onlinecite{2004_senthil}}.
%%%%%%%%%%%%%%%%%%%%%%%%%%%%%%%%%%%%%%%%%%%%%%%%%%%%%%%%%%%%%%%%%%%%%%%%
\subsection{Extension to Other Lattices}
\label{other_lattices}

Having fleshed out the details of the calculation for the case of rectangular lattice, we now briefly point out the extension of this theory to other two dimensional lattices, where similar transitions may arise. To this end, we note that apart from stabilizing the respective phases, the information about the lattice structure enters the analysis in the following way: (1) The point group symmetry of the lattice dictates the transformation of the spinon and only terms invariant under these symmetry transformations are allowed in the action; (2) The vison PSG depends on the point group symmetry of the dual lattice, the latter being related to that of the original lattice.

For the square lattice, repeating the above calculation one finds that there are 2 soft vison modes and the vison action is given by a $XY$ model like Eq. \ref{rect_soft_spin_LGW}, but now with an eight-fold symmetry breaking term $(\cos[8\phi])$. Preliminary analysis suggests that the structure of the theory remains unchanged except for the fact that we now have {\em quadrupled} instanton operator which is connected to the fact that on the square lattice the VBS state has 4-fold degeneracy. 

 Another interesting case is that of the anisotropic triangular lattice, where a similar transition may occur. Starykh {\em et al.} showed that if one takes spin-$1/2$ chains with both nearest neighbour and next nearest neighbour antiferromagnetic interactions and then couple several such chains antiferromagnetically so as to create an anisotropic triangular lattice, then such a microscopic Hamiltonian allows Spiral and dimer phases in proximity to each other \cite{2007_starykh}. It is interesting to ask whether this system allows a direct transition, and if so, what is the critical field theory? We can extend our present analysis to answer the question of structure of the critical theory. Once again preliminary analysis \cite{2007_subhro} suggests that there are four vison modes and the minimal vison action has $O(2)\times O(2)$ symmetry.

Extensions to other lattices may be made similarly. Finally, it is also interesting to note that for analogous cases in 1 D spin chains the visons will always proliferate (since in $(1+1)$ D the $Z_2$ gauge theory is always in a confining phase) leading to dimerization. This helps us to see why the spin-$1/2$ $J_1-J_2$ chain is dimerized in the regime where classical analysis predicts spiral order.

%%%%%%%%%%%%%%%%%%%%%%%%%%%%%%%%%%%%%%%%%%%%%%%%%%%%%%%%%%%%%%%%%%%%%%%%%
\section{Conclusion} 
\label{sec5}
In this paper we have outlined the field theory for a direct second order quantum phase transition between a spiral state and a VBS in context of a concrete spin-1/2 lattice model in two spatial dimensions. This is potentially a new example of deconfined quantum criticality. 

The continuum theory given Eq. \ref{05_final_continum_field_theory} is an anisotropic version of the non-compact $CP^3$ model ($NCCP^3$) and belongs to the general family of $NCCP^{N-1}$ critical theories ($N$ denotes the number of matter components). Anisotropic $NCCP^1$ describes the transition between collinear N\'{e}el and VBS phases in easy-plane spin-$1/2$ $2D$-antiferromagnets {\cite{2004_senthil}}, while anisotropic $NCCP^2$ describes the continuous transition between spin-nematic and VBS in spin-$1$ $2D$-antiferromagnets {\cite{2007_grover}}. Similar to the these other field theories describing the deconfined quantum critical points, the present one is also a strongly coupled theory in (2+1) dimensions. Detailed analysis of such theories require numerical investigation.  
 
%%%%%%%%%%%%%%%%%%%%%%%%%%%%%%%%%%%%%%%%%%%%%%%%%%%%%%%%%%%%%%%%%%%%%%%%%

\begin{acknowledgements}

The author acknowledges T. Senthil for extensive discussion and related collaboration. S. Banerjee, H. R. Krishnamurthy, R. Moessner, S. Mukerjee and D. Sen are thanked for useful discussion.  
\end{acknowledgements}

%%%%%%%%%%%%%%%%%%%%%%%%%%%%%%%%%%%%%%%%%%%%%%%%%%%%%%%%%%%%%%%%%%%%%%%%%%
\appendix

%------------------------------------------------
%%%%%%%%%%%%%%%%%%%%%%%%%%%%%%%%%%%%%%%%%%%%%%%%%%%%%%%%%%%%%%%%%%%%%%%%%%%%%%%%%%%%%%%%%%%%%%%%
\section{Derivation of the Effective Action}
\label{one_appen}

Here we give an alternate way of arriving at the effective action given by Eq. \ref{spinon_action}. The lattice action written in terms of the spiral order parameters has a global $\mathbb{SU}(2)\times\mathbb{U}(1)$ symmetry. The most general action consistent with this symmetry is introduced by the following partition function:
\begin{align}
\mathcal{Z}_{\bf n}&=\int\prod_{\bf i}d{\bf n}_{1i}d{\bf n}_{2i}\ \exp{(-\mathcal{S_{\bf n}}-\mathcal{S}_B)}\\
\mathcal{S}_{\bf n}&=-\tilde{J}\sum_{\langle ij\rangle}\left({\bf n}_{1i}\cdot{\bf n}_{1j}+{\bf n}_{2i}\cdot{\bf n}_{2j}\right)
\end{align}
where $\mathcal{S}_B$ denotes the contribution of the Berry phase term. Using the spinon parametrization as given by Eq. \ref{05_n_z}, this becomes
\begin{align}
\nonumber
\mathcal{Z}_z=\int\prod_idz_{i\alpha}dz^*_{i\alpha}\delta(\vert z_{i\alpha}\vert^2-1)\exp(-\tilde{\mathcal{S}}_z-\mathcal{S}_B),\\
\end{align}
\begin{align}
\nonumber
\tilde{\mathcal{S}}_z=-\tilde{J}_1\sum_{\langle ij\rangle} \vert{\bf z_i^\dagger\cdot z_j\vert^2}-\tilde{J}_2\sum_{\langle ij\rangle}\left({\bf z_i^\dagger\cdot z_j}-{\bf z_j^\dagger\cdot z_i}\right)^2\\
\end{align}
Introducing auxiliary field $L_{ij}$ on the links and performing the Hubbard-Stratonovich decoupling of the first 4-spinon term of the spinon action above, we get
\begin{widetext}
\begin{align}
\mathcal{Z}_z'&=\int\prod_idz_{i\alpha}dz^*_{i\alpha}\ dL_{ij}dL_{ij}^*\ \delta(\vert z_{i\alpha} \vert^2-1)\exp(-\mathcal{S}_z'-\mathcal{S}_B),\\
\label{05_appen_decouple}
\mathcal{S}_z'&=\frac{1}{2\tilde{J}_1}\sum_{\langle ij\rangle}\vert L_{ij}\vert^2-\sum_{\langle ij\rangle}\left(L_{ij}{\bf z_i^\dagger\cdot z_j}+h.c.\right)-\tilde{J}_2\sum_{\langle ij\rangle}\left({\bf z_i^\dagger\cdot z_j}-{\bf z_j^\dagger\cdot z_i}\right)^2,
\end{align}
\end{widetext}
Now following the standard slave particle saddle point treatment {\cite{2000_senthil}} we approximate the auxiliary field by  an uniform saddle point value $L_{ij}=L_0\ (\forall ij)$. The saddle point respects the global symmetries $\mathbb{SU}(2)\times \mathbb{U}(1)$, but it does not obey the local $Z_2$ gauge symmetry of the spinons. The fluctuations of $L_{ij}$ about the saddle point involves both amplitude and the phase fluctuations. It is expected that the fluctuations that restore the $Z_2$ gauge symmetry of the spinons are energetically cheapest. Thus considering this class of fluctuations around the saddle point which re-implements this gauge symmetry, we write
\begin{align}
\label{05_appen_fluctuations}
L_{ij}=L_0\sigma_{ij},
\end{align}
where $\sigma_{ij}=\pm 1$ is a $Z_2$ gauge field defined on the space-time lattice. Using this in Eq. \ref{05_appen_decouple} we get back Eq. \ref{spinon_action}.
%%%%%%%%%%%%%%%%%%%%%%%%%%%%%%%%%%%%%%%%%%%%%%%%%%%%%%%%%%%%%%%%%%%%%%%%%%%%%%%%%%%%%%%%%%%%%%%%
\section{Odd Ising gauge theory, visons and VBS order parameter}
\label{two_appen}

We now consider the massive amplitude fluctuations of $L_{ij}$ in action given by Eq. {\ref{05_appen_decouple}}. About the saddle point let us replace the fluctuations (Eq. \ref{05_appen_fluctuations}) by:
\begin{align}
L_{ij}=(L_0+l_{ij})\sigma_{ij},
\end{align}
where $l_{ij}$ are the amplitude fluctuations of auxiliary field. Putting this  in Eq. \ref{05_appen_decouple} we have,
\begin{align}
\nonumber
\mathcal{S}_z'=&\frac{1}{2\tilde{J}_1}\sum_{\langle ij\rangle}\vert (L_0+l_{ij})\vert^2-\sum_{\langle ij\rangle}\left(\sigma_{ij}(L_0+l_{ij}){\bf z_i^\dagger\cdot z_j}+h.c.\right)\\
&-\tilde{J}_2\sum_{\langle ij\rangle}\left({\bf z_i^\dagger\cdot z_j}-{\bf z_j^\dagger\cdot z_i}\right)^2
\end{align}
Finally integrating out the spinons (we can set $\tilde{J_2}=0$ for this leading order calculation) we generate several terms with the leading order term being :
\begin{align}
\sum_{i,\eta}L^3_0l_{i,i+\eta}(\sigma_{i,i+\eta}\sigma_{i+\eta,i+\eta+\tau}\sigma_{i+\eta+\tau,i+\tau}\sigma_{i+\tau,i}),
\end{align}
where the product of the gauge fields is over the time-like plaquette associated with space-like $l_{i,i+\eta}$ ($\eta$ being the displacement to the nearest space-like neighbour). This coupling between $l_{ij}$ and $\sigma$s result to
\begin{align}
\langle l_{i,i+\eta}\rangle\sim \langle \sigma_{i,i+\eta}\sigma_{i+\eta,i+\eta+\tau}\sigma_{i+\eta+\tau,i+\tau}\sigma_{i+\tau,i}\rangle
\end{align}
or in the time continuum limit of the Ising gauge theory we have:
\begin{align}
\label{eqqq}
\langle l_{i,i+\eta}\rangle\sim\langle \sigma^x_{i,i+\eta}\rangle
\end{align}

At this point we note that the spinon parametrization of Eq. \ref{05_n_z} implies that the VBS order parameter
\begin{align}
\langle{\bf S_i\cdot S_{i+\eta}}\rangle\sim\langle\vert {\bf z_i^\dagger\cdot z_{i+\eta}}\vert^2\rangle
\end{align}
Then from Eq. \ref{05_appen_decouple} we find that ${\bf z_i^\dagger\cdot z_{i+\eta}}\sim L_{\bf i,i+\eta}$. 
Thus we have:
\begin{align}
\langle{\bf S_i\cdot S_{i+\eta}}\rangle\sim L_{\bf i,i+\eta}^2
\end{align}
Thus to the leading order:
\begin{align}
\langle{\bf S_i\cdot S_{i+\eta}}\rangle\sim L_0l_{{\bf i,i+\eta}}.
\end{align}
Using Eq. \ref{eqqq} we finally get
\begin{align}
\label{vbs_finally}
\langle{\bf S_i\cdot S_{i+\eta}}\rangle\sim \langle\sigma_{i,i+\eta}^x\rangle
\end{align}
We now use two well-known results from the duality between Ising gauge theory and Ising model in $(2+1)$ dimensions (we have taken the time continuum limit as before). These are:
\begin{align}
\label{relation1}
\sigma^{x}_{ij}=\rho^z_a\rho^z_b
\end{align}
and 
\begin{align}
\label{relation2}
\mathcal{F}_\Box=\prod_\Box \sigma^z_{ij}=\rho^x_a,
\end{align}
where, in Eq. \ref{relation1}, the link $(ij)$ is on the direct lattice and $(a,b)$ are sites on the dual lattice such that the link $(ab)$ cross $(ij)$. $\rho^\alpha_a$ are the dual Ising spins as defined below Eq. \ref{05_fftfim_ham} in the main paper. In Eq. \ref{relation2},  $\mathcal{F}_\Box$, as defined in the main paper, is the Ising gauge flux through the plaquette of the direct lattice with the dual lattice point $(a)$ at the centre. The implications of Eqs. \ref{relation1} and \ref{relation2} are straight forward. Eq. \ref{relation2} states that $\rho^z_a$ flips the Ising flux and indeed is the vison creation operator. On the other hand, when combined with Eqs. \ref{vbs_finally}, Eq. \ref{relation1} shows that the VBS order parameter is indeed bilinear in the vison creation operators
\begin{align}
\langle{\bf S_i\cdot S_{j}}\rangle\sim \langle\rho^z_a\rho^z_b\rangle.
\end{align}
This completes our derivation for the VBS order parameter. We also note that there are several alternative ways to derive the same result. An example is to start from the fermionic description of spinons as done by Senthil {\em et al.} {\cite{2000_senthil}}.

%%%%%%%%%%%%%%%%%%%%%%%%%%%%%%%%%%%%%%%%%%%%%%%%%%%%%%%%%%%%%%%%%%%%%%%%%%%%%%%%%%%%%%%%%%%%%%%%%
\section{The critical field theory}
\label{three_appen}

The critical theory given by Eq. \ref{the_theory} where different terms are given by Eqs. \ref{spinon_action}, \ref{rect_soft_spin_LGW} and \ref{chern_simmonsterm} respectively. Now the first term in the vison action in Eq. \ref{rect_soft_spin_LGW} may be written as,
\begin{align}
\mu_{ab}\cos{(\phi_a-\phi_b)}=\cos{(\phi_a-\phi_b+\frac{\pi}{2}(1-\mu_{ab}))}.
\end{align}
Within Villain approximation{\cite{1977_villain}}, the cosine term becomes (we henceforth use the notation $\Delta\phi_{ab}=\phi_a-\phi_b$)
\begin{widetext}
\begin{align}
e^{t_v \cos{(\Delta\phi_{ab}+\frac{\pi}{2}(1-\mu_{ab}))}}\approx\sum_{m_{ab}=-\infty}^{\infty}e^{-\frac{t_v}{2}(\Delta\phi_{ab}+\frac{\pi}{2}(1-\mu_{ab})-2\pi m_{ab})^2},
\end{align}
where $m_{ab}$ is an integer defined on the links of the dual lattice. The partition function then written as
\begin{align}
\mathcal{Z}_C=\sum_{\{\sigma_{ij}\}}\sum_{\{\mu_{ab}\}}\sum_{\{m_{ab}\}}\int\prod_id^2{\bf z_i}\ \delta(\vert{\bf z_i}\vert^2-1)\prod_ad{\phi_a}\ \ e^{-\mathcal{S}_z-\mathcal{S}_v-\mathcal{S}_{CS}}
\end{align}
where,
\begin{align}
\mathcal{S}_v=\frac{t_v}{2}\sum_{\langle a,b\rangle} \left(\Delta\phi_{ab}+\frac{\pi}{2}(1-\mu_{ab})\right)^2,
\end{align}
while $\mathcal{S}_z$ and $\mathcal{S}_{CS}$ remain unchanged. Introducing the vison current field $J_{ab}$ on the links of the dual lattice through Hubbard-Stratonovich  decoupling of $\left(\Delta\phi_{ab}+\frac{\pi}{2}(1-\mu_{ab})\right)^2$ we get:
\begin{align}
\mathcal{Z}_C=\sum_{\{\sigma_{ij}\}}\sum_{\{\mu_{ab}\}}\sum_{\{J_{ab}\}}\int\prod_id^2{\bf z_i}\ \delta(\vert{\bf z_i}\vert^2-1)\prod_ad{\phi_a}\ \ e^{-\mathcal{S}_z-\mathcal{S}_v-\mathcal{S}_{CS}}
\label{05_mu_free_sum_z}
\end{align}
where, 
\begin{align}
\mathcal{S}_v=\sum_{\langle ab\rangle} \frac{J_{ab}^2}{2t_v}+iJ_{ab}\left(\Delta\phi_{ab}+\frac{\pi}{2}(1-\mu_{ab})\right)
\end{align}
Regrouping the terms in the critical action, $\mathcal{S}_C$, we get:
\begin{align}
\mathcal{S}_C=\mathcal{S}_z+\sum_{\langle ab\rangle} \frac{J_{ab}^2}{2t_v}+iJ_{ab}\left(\Delta\phi_{ab}\right)+\frac{i\pi}{2}\sum_{\langle ab\rangle} (1-\mu_{ab})\left(J_{ab}-\frac{1-\prod_\Box\sigma_{ij}}{2}\right)
\end{align}
\end{widetext}
The sum over $\mu_{ab}$ can now be freely performed over each link of the dual lattice independently. This yields the constraint  %The resultant partition function has the following form:
%\begin{align}
%\label{05_mu_constraint_z}
%\end{align}
%with the constraint
\begin{align}
\frac{1-\prod_\Box\sigma_{ij}}{2}=J_{ab}+2n_{ab},
\label{05_muj_constraint}
\end{align}
(where $n_{ab}$ is an integer) or,
\begin{align}
\label{current_condition}
(-1)^{j_{ab}}=\prod_{\Box}\sigma_{ij}
\end{align}
and $\mathcal{S}_v$ is given by:
\begin{align}
\mathcal{S}_v=\sum_{\langle ab\rangle} \frac{J_{ab}^2}{2t_v}+iJ_{ab}\left(\Delta\phi_{ab}\right)
\end{align}
Finally integrating over the vison fields $\phi_a$ gives a further constraint on vison current fields $J_{ab}$
\begin{align}
\nabla_b J_{ab}=0
\end{align}

%%%%%%%%%%%%%%%%%%%%%%%%%%%%%%%%%%%%%%%%%%%%%%%%%%%%%%%%%%
\section{Breaking $\mathbb{SU}(2)$ to $\mathbb{U}(1)$}
\label{four_appen}

To find the correct symmetry allowed term that breaks the second $\mathbb{SU}(2)$ to $\mathbb{U}(1)$ we consider the transformation of the ${\bf \chi}$ fields and the ${\bf a_\mu}$ fields under various symmetries of the microscopic Hamiltonian. This may be obtained in several steps, due to the various transformations done to the microscopic Hamiltonian. A fruitful starting point is the transformations of the spinon fields ${\bf z}$ given by Eqs. \ref{spin_sym_trans} and \ref{ref_sym_trans}. We also use the same notation to denote the various transformations. 

The transformations of the spinons immediately imply the transformations for ${\boldsymbol{\nu}}$. The transformation under spin rotation is given by 
\begin{align}
\label{05_nu_trans}
{\nu_\alpha}\rightarrow \tilde{V}_{\alpha\beta}\nu_{\beta}
\end{align}
where $\tilde{V}$ is a superposition of the generators of the first $\mathbb{SU}(2)$. For the rest of the transformations we have
\begin{align}
\nonumber
\mathcal{T}_{\bf a}\ &:\ \ \nu_\alpha\rightarrow \left(e^{-i\frac{\bf Q\cdot a}{2}\tau_y}\right)_{\alpha\beta}\nu_\beta\\
\nonumber
\mathcal{R}_s,\mathcal{I}\ &:\ \ \nu_\alpha\rightarrow -i\left(\tau_x\mu_y\right)_{\alpha\beta}\nu_{\beta}\\
\mathcal{T}\ &:\ \ \nu_\alpha\rightarrow \left(\tau_x\right)_{\alpha\beta}\nu_\beta
\end{align}
Since the gauge fields are minimally coupled to the spinon fields their transformation and also the transformation of $\zeta$ fields are naturally implied by the transformation of the spinon fields. They are given by
\begin{align}
\nonumber
\mathcal{T}_{\bf a}\ &:\ \ \left\{\begin{array}{l}
a_\mu\rightarrow a_{\mu}\\
\zeta\rightarrow \zeta
\end{array}\right.\\
\nonumber
\mathcal{R}_x\ &:\ \ \left\{\begin{array}{l}
a_x\rightarrow a_x,a_y\rightarrow-a_y,a_\tau\rightarrow-a_\tau\\
\zeta\rightarrow-\zeta
\end{array}\right.\\
\nonumber
\mathcal{R}_y\ &:\ \ \left\{\begin{array}{l}
a_x\rightarrow -a_x,a_y\rightarrow a_y,a_\tau\rightarrow-a_\tau\\
\zeta\rightarrow-\zeta
\end{array}\right.\\
\nonumber
\mathcal{I}\ &:\ \ \left\{\begin{array}{l}
a_x\rightarrow -a_x,a_y\rightarrow -a_y,a_\tau\rightarrow-a_\tau\\
\zeta\rightarrow-\zeta
\end{array}\right.\\
\mathcal{T}\ &:\ \ \left\{\begin{array}{l}
a_\mu\rightarrow a_\mu\\
\zeta\rightarrow \zeta
\end{array}\right.
\end{align}
Combining all these, finally we have the transformation for the $\chi$ fields given by:
\begin{align}
\label{trans_chi}
\nonumber
\mathcal{T}_{\bf a}\ &:\ \ \chi_\alpha\rightarrow \left(e^{-i\frac{\bf Q\cdot a}{2}\tau_y}\right)_{\alpha\beta}\chi_\beta\\
\nonumber
\mathcal{R}_s, \mathcal{I}\ &:\ \ \chi_\alpha\rightarrow -i\left(\tau_x\mu_y\right)_{\alpha\beta}\chi^*_\beta\\
\mathcal{T}\ &:\ \ \chi_\alpha\rightarrow \left(\tau_x\right)_{\alpha\beta}\chi^*_\beta
\end{align}
Thus to break the second ${\mathbb{SU}(2)}$ to $\mathbb{U}(1)$ (refer Eq. \ref{eq_su2Xsu2}) we may consider a generic term (It is interesting to note that $\tau_y$ generates translation for the ${\bf \chi}$ fields) 
\begin{align}
\left({\bf \boldsymbol\chi^*\cdot\tau_y\cdot\boldsymbol\chi}\right)^n,
\end{align}
where $n$ is a positive integer. Clearly such a term breaks the second ${\mathbb{SU}(2)}$ to $\mathbb{U}(1)$ and hence produces the correct symmetry of the microscopic Hamiltonian. To determine the lowest value of $n$ consistent with various discrete symmetries we consider the transformation of this term under various symmetries as before:

\begin{align}
\nonumber
\mathcal{T}_{\bf a}\ &:\ \ \left({\bf \boldsymbol\chi^*\cdot\tau_y\cdot\boldsymbol\chi}\right)\rightarrow \left({\bf \boldsymbol\chi^*\cdot\tau_y\cdot\boldsymbol\chi}\right) \\
\nonumber
\mathcal{R}_s, \mathcal{I}\ &:\ \ \left({\bf \boldsymbol\chi^*\cdot\tau_y\cdot\boldsymbol\chi}\right)\rightarrow \left({\bf \boldsymbol\chi^*\cdot\tau_y\cdot\boldsymbol\chi}\right)\\ 
\mathcal{T}\ &:\ \ \left({\bf \boldsymbol\chi^*\cdot\tau_y\cdot\boldsymbol\chi}\right)\rightarrow -\left({\bf \boldsymbol\chi^*\cdot\tau_y\cdot\boldsymbol\chi}\right)
\end{align}

Thus the linear term $(n=1)$ is forbidden by by time reversal symmetry. Thus the lowest order term allowed by all the microscopic symmetries is for $n=2$, {\em i.e.}
\begin{align}
\left({\bf \boldsymbol\chi^*\cdot\Lambda_2\cdot\boldsymbol\chi}\right)^2,
\end{align}

%%%%%%%%%%%%%%%%%%%%%%%%%%%%%%%%%%%%%%%%%%%%%%%%%%%%%%%%%%%%%%%%%%%%%%%%%%%%%%%%%%%%%%%%%%%%%%%%
%\end{widetext}

%%%%%%%%%%%%%%%%%%%%%%%%%%%%%%%%%%%%%%%%%%%%%%%%%%%%%%%%%%%%%%


\begin{thebibliography}{11}
\bibitem{2008_sachdev} S. Sachdev, Nature Physics {\bf 4}, 173 (2008).
\bibitem{2004_senthil} T. Senthil, A. Vishwanath, L. Balents, S. Sachdev and M. P. A. Fisher, Science {\bf 303}, 1490 (2004); Phys Rev.  {\bf B 70}, 144407 (2004); J. Phys. Soc. Jap. {\bf 74}, Suppl. 1 (2005).
\bibitem{2005_ghaemi} P. Ghaemi and T. Senthil, Phys Rev. {\bf B 73}, 054415 (2006)
\bibitem{1994_chubukov} A. V. Chubukov, S. Sachdev, and T. Senthil, Nucl. Phys. {\bf B 426}, 601 (1994); Phys. Rev. Lett. {\bf 72}, 2089 (1994).
\bibitem{1991_read} N. Read and S. Sachdev, Phys. Rev. Lett. {\bf 66}, 1773 (1991); S. Sachdev, Phys. Rev. {\bf B 45}, 12377 (1992).
\bibitem{1984_kawamura} H. Kawamura and S. Miyashita, J. Phys. Soc. Jap. {\bf 53}, 4138 (1984).
\bibitem{1976_ma} S. K. Ma, Mordern Theory of critical Phenomena, Perseus Books (1976).
\bibitem{1996_leung} P. W. Leung and Ngar-wing Lam, Phys. Rev. {\bf B 53}, 2213 (1996).
\bibitem{1992_nomura} K. Okamoto and K. Nomura, Phys. Lett. {\bf A 169}, 433 (1992).
\bibitem{2004_capriotti} L. Capriotti, D. J. Scalapino, and Steven R. White, Phys. Rev. Lett. {\bf 93}, 177004 (2004). 
\bibitem{1985_tung} Wu-Ki. Tung, {\it Group Theory in Physics}, (World Scientific, Singapore) (1985).  
\bibitem{2003_sachdev_2} S. Sachdev, Rev. Mod. Phys. {\bf 75}, 913 (2003); Ann. Phys. {\bf 303}, 226 (2003).
\bibitem{1979_mermin}N. D. Mermin, Rev. Mod. Phys. {\bf 51}, 591 (1979).
\bibitem{1990_angelucci} A. Angelucci, Int. J. Mod. Phys. {\bf 4}, 569 (1990); Phys. Rev. {\bf B 45}, 5387 (1992).
\bibitem{2001_goldstein} H. Goldstein, C. P. Poole, and John L. Safko, {\it Classical Mechanics}(3rd Ed.), (Addison Wesley) (2001).
\bibitem{2000_senthil} T. Senthil and M. P. A. Fisher, Phys. Rev. {\bf B 62},7850 (2000).
\bibitem{1993_lammert} P. E. Lammert, D. S. Rokshar and J. Tonar, Phys. Rev. Lett. {\bf 70}, 1650 (1993).
\bibitem{1989_dombre} T. Dombre and N. Read, Phys. Rev. {\bf B 39}, 6797 (1989).
\bibitem{1993_diptiman} S. Rao and D. Sen, Phys. Rev. {\bf B 48}, 12763 (1993).
\bibitem{2000_sachdev} S. Sachdev and M. Vojta, J. Phys. Soc. Jap. {\bf 69},Supplement B, 1 (2000).
\bibitem{1971_wegner} F. Wegner, J. Math. Phys. {\bf 12}, 2259 (1971).
\bibitem{2001_moessner} R. Moessner, S. L. Sondhi and E. Fradkin,   Phys. Rev. {\bf B 65}, 024504 (2001).
\bibitem{1979_kogut} J. B. Kogut, Rev. Mod. Phys. {\bf 51}, 659 (1979).
\bibitem{1980_savit} R. Savit, Rev. Mod. Phys. {\bf 52}, 453 (1980).
\bibitem{1991_jalabert} R. A. Jalabert and S. Sachdev, Phys. Rev. {\bf B 44}, 686 (1991).
\bibitem{1984_blanckstein} D. Blankschtein, M. Ma and A. N. Berker, Phys. Rev. {\bf B 30}, 1362 (1984).
\bibitem{2001_lannert} C. lannert, T. Senthil and M. P. A. Fisher, Phys. Rev. {\bf B 63}, 134510 (2001).
\bibitem{2002_wen} X. G. Wen, Phys. Rev. {\bf B 65}, 165113 (2002).
\bibitem{1977_villain} J. Villain, J. Phys. {\bf C 10}, 1717 (1977).
\bibitem{2009_xu} C. Xu and S. Sachdev, Phys. Rev. {\bf B 79}, 064405 (2009).
\bibitem{2007_grover} T. Grover and T. Senthil, Phys. Rev. Lett. {\bf 98}, 247202 (2007).
\bibitem{1987_polyakov} A. M. Polyakov, {\it Gauge Fields and Strings}, (Harwood, Newyork) (1987).
\bibitem{1979_fradkin} E. Fradkin and S. H. Shenker, Phys. Rev. {\bf D 19}, 3682 (1979).
\bibitem{1990_murthy} G. Murthy and S. Sachdev, Nucl. Phys. {\bf B 344}, 557 (1990).
\bibitem{2007_sandvik} A. W. Sandvik, Phys. Rev. Lett. {\bf 98}, 227202 (2007).
\bibitem{2008_kou} S.-P. Kou, M. Levin, and X.-G. Wen, Phys. Rev. {\bf B 78}, 155134 (2008).
\bibitem{2007_starykh} O. A. Starykh and L. Balents, Phys. Rev. Lett. 98, 077205 (2007). 
\bibitem{2007_subhro} S. Bhattacharjee, MS Report, Indian Institute of Science, Bangalore (2007) (unpublished).
\end{thebibliography}
\end{document}